# SVM based on personal identification system using Electrocardiograms

Emna Rabhi[#1], Zied Lachiri[*2]

[#] University of Tunis El Manar
National School of Engineers of Tunis, LR11ES17 Signal, Image
and Information Technology laboratory, BP. 37 Le Belvédère, 1002, Tunis, Tunisia
[1] emna.rabhi@gmail.com

[*] National Institute of Applied Science and Technology
BP. 676 centre urbain cedex Tunis, Tunisia
[2] zied.lachiri@enit.rnu.tn

*Abstract*— This paper presents a new algorithm for personal identification from their Electrocardiograms (ECG) which is based on morphological descriptors and Hermite Polynomials Expansion coefficients (HPEc). After preprocessing, we extracted ten morphological descriptors which were divided into homogeneous groups (amplitude, surface interval and slope) and we extracted sixty Hermite Polynomials Expansion coefficients (HPEc) from each heartbeat. For the classification, we employed a binary Support Vector Machines with Gaussian kernel and we adopted a particular strategy: we first classified groups of morphological descriptors separately then we combined them in one system. On the other hand, we classified the Hermite Polynomials Expansion coefficients apart and we associated them with all groups of morphological descriptors in a single system in order to improve overall performance. We tested our algorithm on 18 different healthy signals of the MIT_BIH database. The analysis of different groups separately showed that the best recognition performance is 96.45% for all morphological descriptors and the results of experiments showed that the proposed hybrid approach has led to an overall maximum of 98.97%.

I. INTRODUCTION

The conventional means of identification such as passwords, secret codes and personal identification numbers (PIN) can be easily shared, observed, stolen or forgotten. However, a possible alternative in determining the identity of users is to use biometrics. Recognition of the person biometric refers to the process of automatic recognition of a person using behaviours (approach, signature, keystroke, lip movement, handle) or physiological traits (face, voice, iris, fingerprint, hand geometry, Electroencephalogram EEG, ECG electrocardiogram, ear shape, body odor). In recent decades, many of these biometric modalities were studied biometric modalities (fingerprint, iris, voice, face) and are still being examined [1].

In the recent past, various studies have been conducted for the use of biometrics in the ECG [10].

Biel and al.[5] demonstrated for the first time the feasibility of using ECG signals for human identification. A SIEMENS ECG apparatus was used to record cardio-vascular signals and at the same time to extract a set of temporal, amplitude and slope features, then, a simple analysis of correlation matrix is used to reduce the dimensionality of features. Finally, a SIMCA model based on principal components analysis PCA is employed to classification, but the major drawback of the method was the lack of automatic recognition, since specific apparatus was used for feature extraction.

Ming Li [7], [8] used an algorithm based on time and cepstral information applied to 18 healthy individuals. After preprocessing, and extraction features, the temporal parameters were modeled by SVM with a linear kernel and cepstral parameters were modeled by a GMM then he made a weighted sum fusion for improved result which reaches 98.3%. Issues of these studies are mainly: the feature extraction of ECG and physiological changes of the heart.

In summary, the main drawback of the previous works is the low accuracy in the automatic detection of fiducial points as there is no universally acknowledged rule for defining exactly where the wave boundaries lie [4]. In this paper, we first generalize the existing literature by applying the same analytical characteristics, which attributes the temporal distance and amplitude. To improve the identification accuracy, we combined morphological descriptors with the coefficients of the Hermite polynomials, so at the end we will have an output vector composed of both 10 morphological descriptors and 60 Hermite Polynomial Expansion coefficients. For the classification, we used Binary Support Vector Machines.

This paper is organized as follows: Sections2 describes the proposed method and illustrates the extraction features and the classification system adopted, Section3 presents the experimental results. Finally the conclusions reached will be the last section of this paper.

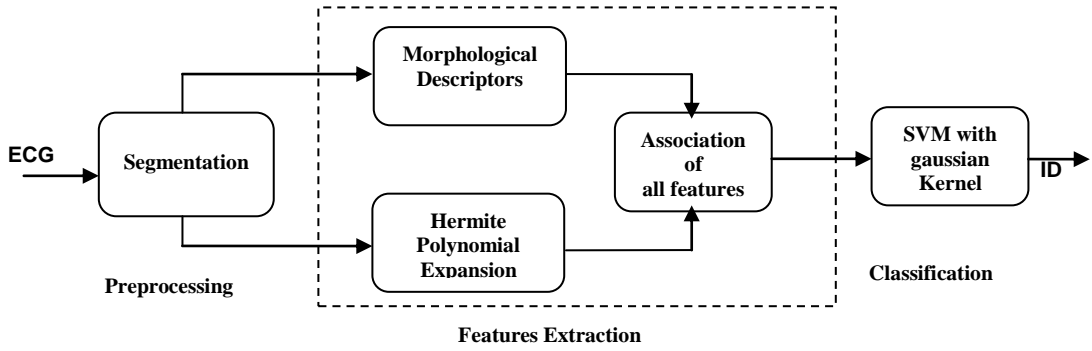

*Fig. 1 Block diagram of the adopted algorithm.*

## II. PROPOSED METHOD

The proposed system consists of three blocks which are preprocessing, features extraction and classification. After preprocessing, ten morphological descriptors are extracted from the each heartbeat of the ECG signals and which were divided into homogenous groups (amplitude, surface, interval and slope). Later, sixty Hermite Polynomials Expansion coefficients (HPE) are extracted from the ECG signal. The groups of morphological descriptors were used separately in the classification and associations were made between these groups. At last, to improve the overall performance of recognition, we tested our system with an input vector formed by all features. The parameter vector contains the 10 morphological descriptors and the 60 coefficients of each heartbeat. We adopted this association in the topology SVM shown in Fig.6.

### A. Morphological descriptors

Preprocessing of the ECG signal consists in sampling the QRS complex and determinig the point of reference for the extremum of this complex (R)[11]. We used the list of reference signals provided by the MIT_BIH database. Each reference point (R), has identified the beginning of the QRS (Q = R - 50 ms) and end (S + R = 100ms) as shown in Fig. 1. Therefore the interval QS is of the order of 150 ms, or 54 samples for a sampling frequency of 360 Hz.

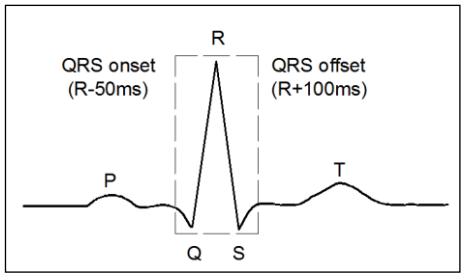

*Fig. 2 Description of QS interval.*

Several morphological descriptors [3] can be extracted from each complex. The ten parameters extracted from each ECG event are described as shown in TABLE I.

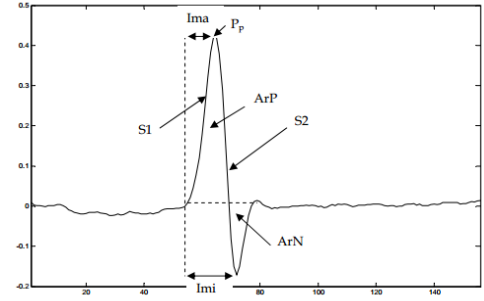

*Fig. 3 The morphological descriptors.*

TABLE I. Description of morphological descriptors

| Type | Description |
|---|---|
| $P_p$ | Maximal amplitude of the positive peak |
| $P_n$ | Maximal amplitude of the negative peak |
| ArP | Surface of the positive samples |
| ArN | Surface of the negative samples |
| Ar | Surface of the QRS (ArP+ArN) |
| No | Nomber of samples with 70 higher amplitude |
| Ima | Time-interval from the QRS complex onset to the maximal positive peak |
| Imi | Time-interval from the QRS complex onset to the maximal negative peak |
| S1 | QRS slope velocity calculated for the time-interval between QRS complex onset and the first peak |
| S2 | QRS slope velocity calculated for the time-interval between the first peak and the second peak |

### B. Hermite polynomials expansion (HPE)

The approximation and estimation of physical and real signals as the cardiac signals are of great importance in the biomedical and especially if the functions used for this estimate have a reduced number of parameters. The Hermite polynomials have been successfully used to describe the ECG signal due to the similarity existing between the forms and functions of Hermite waveforms of the ECG [6]. Each heart beat was normalized and then segmented using standard methods of ECG toolbox [4].

Let a(t) the vector curve modeling each heart beat and L the polynomial order. The Hermite polynomials are represented by:

$$a(t) = \sum_{n=0}^{L-1} c_n \phi_n(t,\delta) \quad t \in [-M, M] \quad (1)$$

with $C_n (n=0...L-1)$ are the Hermite polynomials Expansion coefficients and $\phi_n(t,\delta)$ is the Hermite basis functions which are defined as follows:

$$\phi_n(t,\delta) = \frac{1}{\sqrt{\delta 2^n n! \sqrt{\pi}}} e^{\frac{-t^2}{2\delta^2}} H(t/\delta) \quad (2)$$

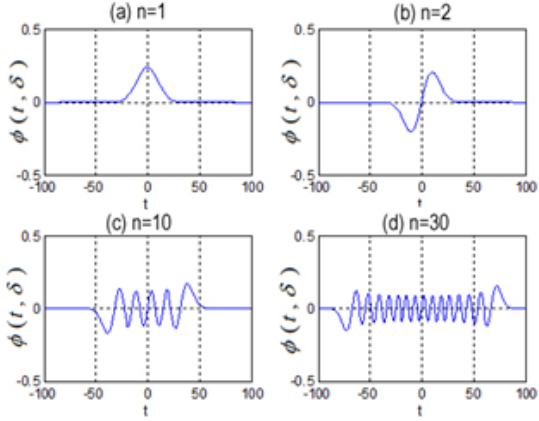

Fig. 4 Four Hermite functions with M= 100 and $\delta = 10$

The Hermite functions $H_n(t)$ are the physicists' Hermite polynomials [6] which are defined recursively by:

$$H_0(t) = 1, \ H_{(1)} = 2t, \ H_{(3)} = t^2 - 1 \quad (3)$$
$$H_n(t) = 2tH_{n-1}(t) - 2(n-1)H_{n-2}(t) \quad (4)$$

We observe from Fig.4 that the higher the order of Hermite basis functions, the higher is its oscillation, which leads to a greater ability to capture morphological details.

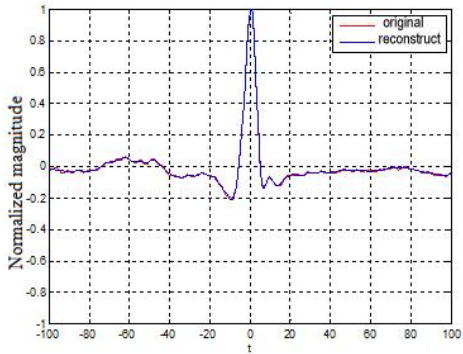

Fig. 5 An original heartbeat and a reconstructed heartbeat by HPE

During the reconstruction of the ECG signal by Hermite Polynomial Expansion, one notices the likeness between an original heartbeat and a reconstructed heartbeat by HPE as shown in Fig.5.

*C. SVM classification*

For the classification, we used Binary Support Vector Machines. In the nonlinear case, the idea is to use a kernel function $K(x_i, x_j)$, where $K(x_i, x_j)$ satisfies the Mercer conditions [12].

In order to find the SVM that better generalization performance, the simplest way consists on using an independent set of the training for the classifier assessments. We adopted approach two-thirds one-third that is to say 2/3 samples were used for training and the remaining 1/3 for testing.

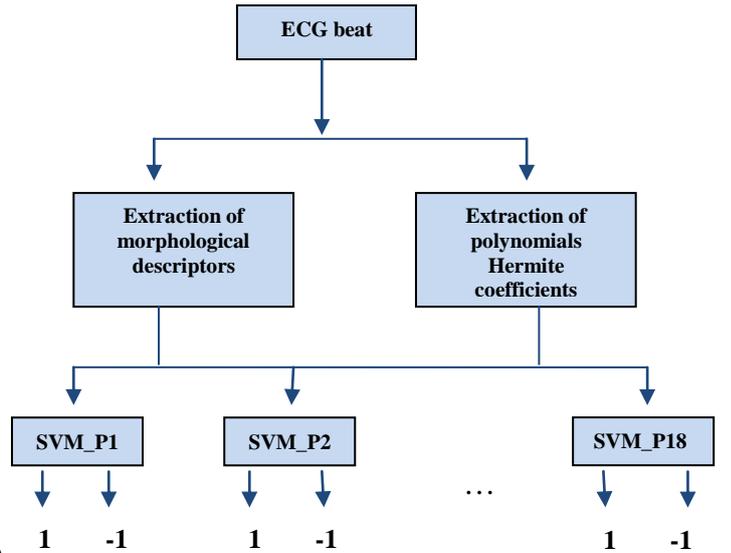

Fig. 6 Classification model by association of all descriptors.

The success of SVMs depends on two tricks which are selecting the best hyperplane and the best kernel function. The best hyperplane is find by making the margin largest. The extension of this concept to a higher dimensional setting using kernel functions to represent a similarity measure on that setting.

We used a polynomial and a Gaussian kernel [2] which are defined as follows:

$$K_{RBF}(x, x') = \exp\left(-\frac{\|x - x'\|^2}{2\sigma^2}\right) \quad (5)$$

$$K_{poly}(x, x') = \left(a * \langle x, x' \rangle + b\right)^d \quad (6)$$

## III. EXPERIMENTAL RESULTS

We tested our approach on the database arrhythmia MIT / BIH [9]. Only 18 signals from different healthy individuals were used for assessments. Each recording lasts about 30 minutes with a sampling frequency of 360 Hz.

TABLE II. Global rate with different features

| Parameters group | Global rate (%) |
|---|---|
| $P_n+P_p$ | 95 |
| ArN+Ar+ArP | 95 |
| No+Ima+Imi | 94.99 |
| S1+S2 | 95.02 |
| $(P_n+P_p)$ and (ArN+Ar+ArP) | 95.91 |
| (ArN+Ar+ArP) and (No+Ima+Imi) | 95.01 |
| (ArN+Ar+ArP) and (S1+S2) | 95 |
| $(P_n+P_p)$ and (No+Ima+Imi) | 94.88 |
| (S1+S2) and (No+Ima+Imi) | 95 |
| $(P_n+P_p)$ and (S1+S2) | 95 |
| $(P_n+P_p)$,(ArN+Ar+ArP),(No+Ima+Imi) and (S1+S2) | 96.45 |
| $C_n(n=1,..,60)$ | 96.33 |
| $(P_n+P_p)$,(ArN+Ar+ArP)(No+Ima+Imi),(S1+S2) and $C_n$ | **98.97** |

For classification, we used the approach two-thirds one-third that is to say 20 min of the data about 2/3 samples were used for training and the remaining 1/3 for testing about the last 10 min. As for the SVM topology, we have chosen two kernels: the Gaussian and polynomial. Experiments were conducted to determine the appropriate values of hyper-parameters, namely the order of the polynomial which varied from 0.25 to 2, the Gaussian width $\sigma$ that varied from 0.25 and 1 and the regularization parameter C ranged between 10 and 1000. The best rates are obtained for SVM model with a Gaussian kernel, a Gaussian width $\sigma = 0.5$ and a regularization parameter C = 1000. We adopted a particular strategy of the parameters choice of binary Support Vector Machines in order to improve the classification performance. Firstly, we used separately four groups of the morphological descriptors: Amplitude (Pp and Pn), Surface (Ar, ArP, ArN), interval (No, Ima, Imi) and slope (S1, S2). Then, we combined them as input parameters of our system. On the other hand, we employed the Hermite Polynomials Expansion coefficients (HPEc) which were calculated with a polynomial order equal to 60, M = 100 and δ = 10. Finally, we associated the HPE coefficients with all groups of morphological descriptors and we obtained an input vector that contains the 60 HPE coefficients and ten morphological descriptors.

The experimental results are illustrated in TABLE II which reports the identification rates obtained for each case. We observed that the groups of the morphological descriptors have yielded a similar identification rate. For example, the rate of identification produced using amplitude group (Pp + Pn) equal to 95% and the rate of interval group is 94.99% .

As figured in table II, the association between all groups of the morphological descriptors has improved the performance result reaching 96.45%.

As it can be seen, the Hermite polynomials Expansion coefficients have achieved a classification rate equal to 96.33%.

The best performance of our system has been attained by combining the all groups of morphological descriptors with the HPE coefficients. Thus, we have obtained an identification rate equal to 98.97%. This result is slightly higher than that given by the M. Li and al. [8] approach (98.26%); which is based on the fusion of Hermite Polynomials Expansion and the cepstral coefficients.

IV. CONCLUSION

In this work, a new algorithm was proposed for biometric identification of individuals. To effectively model the intra cardiac models, we combined the Hermite polynomials expansion coefficients and the morphological parameters through modeling by the SVM with a Gaussian kernel. For further work, we can make the validation of the method on a larger number of people and improve the method by merging it with another technique to use it on ECG signals having different types of pathologies.